\documentclass{article}

    \usepackage[preprint]{neurips_2023}

\usepackage[utf8]{inputenc} 
\usepackage[T1]{fontenc}    
\usepackage{hyperref}       
\usepackage{url}            
\usepackage{booktabs}       
\usepackage{amsfonts}       
\usepackage{nicefrac}       
\usepackage{microtype}      
\usepackage{xcolor}         
\usepackage{tablefootnote}
\usepackage{amsmath, bm}
\usepackage{algorithm}
\usepackage{algpseudocode}
\title{CoMoSpeech: One-Step Speech and Singing Voice Synthesis via Consistency Model}

\usepackage{graphicx}

\author{Zhen Ye\textsuperscript{1}, Wei Xue\textsuperscript{1$\dag$}, Xu Tan\textsuperscript{2}, Jie Chen\textsuperscript{3}, Qifeng Liu\textsuperscript{4,1}, Yike Guo\textsuperscript{1$\dag$} 
\\
\textsuperscript{1}	Hong Kong University of Science and Technology \textsuperscript{2}	Microsoft Research Asia \\
\textsuperscript{3}  Hong Kong Baptist University\\ \textsuperscript{4}	 Hong Kong Institute of Science \& Innovation, Chinese Academy of Sciences
\thanks{
$^\dag$Corresponding authors: Wei Xue \{\small weixue@ust.hk\}, Yike Guo
        \{\small yikeguo@ust.hk\}        
}
}

\begin{document}

\maketitle

\begin{abstract}
Denoising diffusion probabilistic models (DDPMs) have shown promising performance for speech synthesis. However, a large number of iterative steps are required to achieve high sample quality, which restricts the inference speed. Maintaining sample quality while increasing sampling speed has become a challenging task. In this paper, we propose a \textbf{Co}nsistency \textbf{Mo}del-based {Speech} synthesis method, CoMoSpeech, which achieve speech synthesis through a single diffusion sampling step while achieving high audio quality. The consistency constraint is applied to distill a consistency model from a well-designed diffusion-based teacher model, which ultimately yields superior performances in the distilled CoMoSpeech. 
Our experiments show that by generating audio recordings by a single sampling step, the CoMoSpeech achieves an inference speed more than 150 times faster than real-time on a single NVIDIA A100 GPU, which is comparable to FastSpeech2, making diffusion-sampling based speech synthesis truly practical. Meanwhile, objective and subjective evaluations on text-to-speech and singing voice synthesis show that the proposed teacher models yield the best audio quality, and the one-step sampling-based CoMoSpeech achieves the best inference speed with better or comparable audio quality to other conventional multi-step diffusion model baselines. Audio samples and codes are available at \href{https://comospeech.github.io/}{https://comospeech.github.io/}.
\end{abstract}


 


\maketitle

\section{Introduction}

Speech synthesis \cite{tan2021survey}  aims to produce realistic audio of humans and has broadly included text-to-speech (TTS) \cite{taylor2009text,shen2023naturalspeech} and singing voice synthesis (SVS) \cite{nishimura2016singing} tasks due to the increasing applications in human-machine interaction and entertainment. The mainstream of speech synthesis has been dominated by the deep neural network (DNN)-based methods \cite{wang2017tacotron} \cite{kim2021conditional}, and typically a two-stage pipeline is adopted \cite{ren2019fastspeech} \cite{lu2020xiaoicesing}, in which the acoustic model first converts the textual and other controlling information into acoustic features (e.g., mel-spectrogram) and then the vocoder further transforms the acoustic features into audible waveforms. The two-stage pipeline has achieved substantial success since the acoustic features, which are expressed by frames, effectively act as the ``relay'' to alleviate the one-to-many mapping problems (ill-posed or ill-condition problem) \cite{bertero1988ill}   of converting short texts to long audios with a high sampling frequency. 

The quality of the acoustic feature produced by the acoustic model, typically mel-spectrogram, crucially affects the quality of the synthesized speeches. Approaches widely used in the industry, such as Tacotron \cite{wang2017tacotron}, DurIAN \cite{yu2019durian}, and FastSpeech \cite{ren2019fastspeech}, generally adopt the convolutional neural network (CNN) and Transformers to predict the mel-spectrogram from the controlling factor. Diffusion   model  methods have attracted much attention because their potential to produce high-quality samples is well recognized. 
\begin{figure}[!t]
      \centering
      \includegraphics[width=80mm]{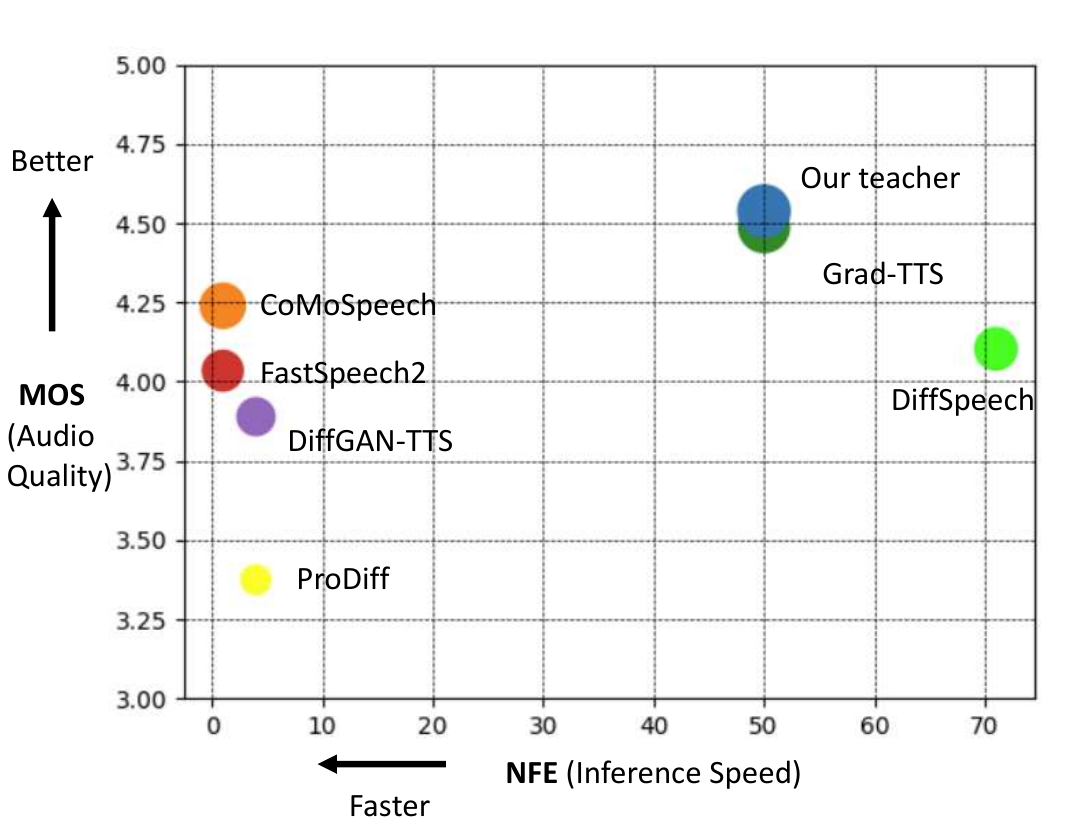}
      \caption{The audio quality and inference speed comparisons of different TTS systems. Details are shown in Table 1, and similar results are obtained for SVS.}
      \label{fig2x}
\end{figure}

A diffusion model  \cite{ho2020denoising}, also named score-based model \cite{song2020score}, is based on two processes, a   diffusion process that gradually perturbs data to noise and a reverse   process  that progressively converts noise back to data. A critical drawback \cite{song2020denoising} \cite{yang2022diffusion} of the diffusion model is that it requires many iterations for the generation.  Several  methods based on the diffusion model have been proposed for acoustic modeling in speech synthesis. Most of these works still have the issue of slow generation speed.

Grad-TTS \cite{popov2021grad} apply the diffusion model for acoustic modeling,  which formulates a stochastic differential equation (SDE) \cite{anderson1982reverse} to gradually transform the noise to the mel-spectrogram and a numerical ODE solver is used for solving reverse SDE \cite{song2020score}. Although yielding high audio quality, the inference speed is low due to the large number of iterations ($10\sim1000$ steps) in the reverse process.  
Prodiff \cite{huang2022prodiff} was further developed to use progressive distillation \cite{salimans2022progressive} to reduce the sampling steps. In \cite{liu2022diffgan}, DiffGAN-TTS adopted an adversarially-trained model to approximate the denoising function for efficient speech synthesis.   In \cite{chen2022resgrad}, the ResGrad uses the diffusion model to estimate the prediction residual between pre-trained FastSpeech2 \cite{ren2020fastspeech} and ground truth.  Apart from normal speaking voice,  recent studies also 
focus on voice with more complex variations in pitch, timing, and expression. For example, Diffsinger \cite{liu2022diffsinger} also shows that a well-designed diffusion model can achieve high quality on synthesized singing voice through one hundred steps of iteration.

From the above discussion, the objectives of speech synthesis are three-fold:
\begin{itemize}
\item  High audio quality: The generative model should accurately express the nuances of speaking voice which contribute to the naturalness and expressiveness of the synthesized audio. Additionally, artefacts and distortions in the generated audio should also be avoided.

\item   Fast inference speed: Real-time applications, including communication, interactive speech and music systems,  require the fast generation speed of audio. When considering making time for other algorithms in an integrated system, simply being faster than real-time is insufficient for speech synthesis.

\item Beyond speech: Instead of the normal speaking voice, more complex modeling of voice on pitch, expression, rhythm, breach control and timbre is required such as singing voice.

\end{itemize}

Although many efforts have been made, due to the mechanism of the denoising diffusion process when performing sampling, the trade-off problem among the synthesized audio quality, model capability and inference speed still exists in TTS and is particularly pronounced in SVS. Existing methods generally seek to alleviate the slow inference problem rather than solve it fundamentally, and their speed is still not comparable to conventional methods without relying on diffusion models such as FastSpeech2 \cite{ren2020fastspeech}. Recently, by expressing the stochastic differential equation (SDE) describing the sampling process as an ordinary differential equation (ODE), and further enforcing the consistency constraint of the ODE   trajectory, the consistency model \cite{song2023consistency}   has been developed, yielding high-quality images with only one sampling step.
However, despite such success in image synthesis, no speech synthesis model based on the consistency model is known so far.
This indicates the potential of designing a consistency model based speech synthesis method to achieve both high-quality synthesis and fast inference speed.

In this paper, we propose \textbf{Co}nsistency \textbf{Mo}del based method for speech synthesis, namely CoMoSpeech, which achieves fast and high-quality audio generation. Our CoMoSpeech is distilled from a pre-trained teacher  model. More specifically, our teacher model leverages the SDE to smoothly transform the  mel-spectrogram  into the Gaussian noise distribution and learn the corresponding score function. After training, 
we utilize the corresponding numerical ODE solvers to construct the teacher denoiser function, which is used for further consistency distillation. Through consistency distillation, our CoMoSpeech is obtained. Ultimately, high-quality audio can be produced by our CoMoSpeech with a  single-step sampling.

We conducted experiments for both TTS and SVS, and the results show that the CoMoSpeech can generate speeches with one sampling step, more than 150 times faster than in real-time. The audio quality evaluation also shows that the CoMoSpeech achieves better or comparable audio quality to other diffusion model methods involving tens to hundreds of iterations (visualized in Figure 1). This makes the speech synthesis based on the diffusion model truly practical for the first time.

\section{Background of Consistency Model}
\label{background}
Now we briefly introduce the consistency model. 
Supposing that we have a data distribution as ${p_\textrm{data}}(\mathbf{x})$. The diffusion model progressively adds the Gaussian noise to diffuse data and then adopts a reverse denoising process to generate samples from noise. For noisy data $\{\mathbf{x}\}_{t=0}^T$ in the diffusion process where $p_0(\mathbf{x})={p_\textrm{data}}(\mathbf{x})$, $p_T(\mathbf{x})$ infinitely close a Gaussian distribution, and $T$ is the time constant, the forward diffusion process can be expressed by a SDE \cite{song2020score} as
\begin{equation}
\mathrm{d} \mathbf{x}=f(\mathbf{x},t)\mathrm{d}t+g(t) \mathrm{d} \mathbf{w},
\label{sde0}
\end{equation}
where $ \mathbf{w}$  is the standard wiener process, $f(\cdot,\cdot)$ and $g(\cdot)$ are drift and diffusion coefficients, respectively. $f(\mathbf{x},t)$  acts as   $f(\mathbf{x},t) = f(t)\mathbf{x}$ in previous work (VP, VE, EDM) \cite{song2020score}, \cite{karras2022elucidating}, thus 
\begin{equation}
\mathrm{d} \mathbf{x}=f(t)\mathbf{x}\mathrm{d}t+g(t) \mathrm{d} \mathbf{w}.
\label{sde}
\end{equation}
A notable property of the above SDE is that it corresponds to a probability flow ODE which indicates the sampling trajectory distribution of SDE at time $t$ \cite{song2020score,karras2022elucidating}, as 
\begin{equation}
\mathrm{d} \mathbf{x}=\left[f(t) \mathbf{x}-\frac{1}{2} g(t)^2 \nabla \log p_t(\mathbf{x})\right] \mathrm{d} t,
\label{ode}
\end{equation}
where ${\nabla }\log {p_t}({\mathbf{x}})$ is the score function of ${p_t(\mathbf{x})}$ \cite{hyvarinen2005estimation}. The probability flow ODE eliminates the stochastic $\mathbf{w}$, thus generating a deterministic sampling trajectory.
\begin{figure*}[!ht]
      \centering
      \includegraphics[width=130mm]{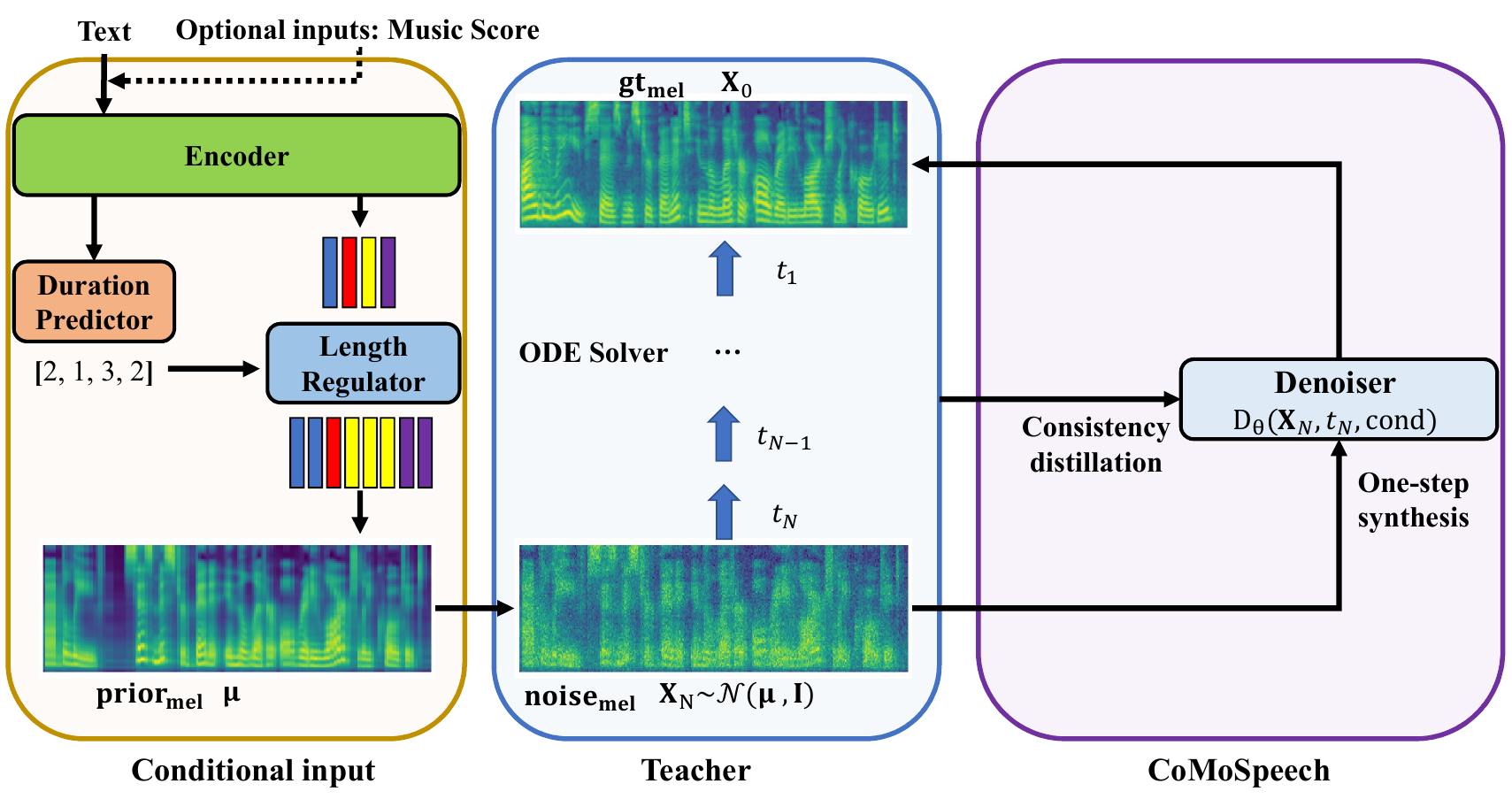}
      \caption{ An illustration of  CoMoSpeech. Our CoMoSpeech distills the multi-step sampling of the teacher model into one step utilizing the consistency constraint.}
      \label{fig1}
\end{figure*}

As long as the score function ${\nabla }\log {p_t}({\mathbf{x}})$ is known, the  probability flow ODE in \eqref{ode} can be used for sampling. Supposing $D(\mathbf{x}_t, t )$ is the ``denoiser'' which denoise the sample $\mathbf{x}_t$ at step $t$, the score function can be obtained by minimizing the denoising error $||D(\mathbf{x}_t, t )-\mathbf{x}||^2$ \cite{karras2022elucidating}, yielding:
\begin{equation}
{\nabla }\log {p}_t({\mathbf{x}}) =(D(\mathbf{x}_t, t )-\mathbf{x}_t) / \sigma_t^2,
\label{score}
\end{equation}
where $\sigma _t^2 = \int {g{{(t)}^2}} dt$. Further, the probability flow ODE based sampling can be performed by first sampling from a noise distribution and then denoising to the true sample by the numerical ODE solver  such as Euler and Heun solvers \cite{song2020score} \cite{karras2022elucidating}. However, the ODE solvers still involve many iterations causing a slow sampling. 

To accelerate sampling \cite{song2023consistency} or minimize the sampling drift  \cite{daras2023consistent},  consistency property has been proposed for diffusion model to impose both:
\begin{equation}
D(\mathbf{x}_t , t ) = D(\mathbf{x}_{t^{\prime}} , t^{\prime} )
\label{s1}
\end{equation}
for any $t$ and $t^{\prime}$, and 
\begin{equation}
D(\mathbf{x}_0 , 0 ) = \mathbf{x}_0.
\label{s2}
\end{equation}
In this way, a consistency model can be obtained, and one-step sampling $D(\mathbf{x}_T, T )$ can be achieved since all points on a sampling trajectory of probability flow ODE is directly linked to the trajectory’s origin. $p_0(\mathbf{x})$. The consistency model can be trained either in isolation or by distilling from a pre-trained diffusion-based teacher model, and the later approach generally produces better performances. Detailed discussions can be referred to \cite{song2023consistency}. In our work, a distillation-based consistency model for speech synthesis, called CoMoSpeech, is proposed below.

\section{CoMoSpeech}
This section presents the proposed CoMoSpeech, a one-step speech synthesis model. The framework of the proposed method is shown in Figure. \ref{fig1}, which has two main stages. The first stage trains a diffusion-based teacher model to produce audios conditioned on the textual (for TTS and SVS) and musical score inputs (for SVS). Then in the second stage, by forcing the consistency property, we obtain the CoMoSpeech from the distillation of the teacher model to finally achieve a one-step inference given the conditional inputs. How to design the teacher model, perform consistency distillation and implement the training and inference will be discussed.

\subsection{Teacher Model}
\label{teacher}
As a blossoming class of generative models, many speech synthesis systems apply diffusion models and generate high-quality audio. However, specific criteria must be met to be the teacher model. First, the model needs to meet the theoretical requirement. As mentioned in Section \ref{background}, we aim to adopt the denoiser to implement the one-step generation, which means this function should point to the clean data instead of noise. In other words, we follow the term in \cite{huang2022prodiff} that our teacher model should be a generator-based rather than a gradient-based method. This restriction requires us to modify the state-of-art model Grad-TTS \cite{popov2021grad} to be our teacher model. We inherited the setting of training and the main architectures. In addition, we also adopt the EDM\cite{karras2022elucidating} as our design choice for the diffusion model to ensure further consistency distillation \cite{song2023consistency}.

Specifically,  we set mel-spectrogram as $\mathbf{x}$ in \eqref{sde} with the schedule $\sigma (t)$ and scaling coefficients in EDM \cite{karras2022elucidating} as $t$ and 1, respectively. Combined with \eqref{score}, our ODE can be formulated as
\begin{equation}
\mathrm{d} \mathbf{x}_t= [(\mathbf{x}_t -  D_{\theta}(\mathbf{x}_t , t,cond))  / t ] \mathrm{d} t,
\end{equation}
where $cond$ is the conditional input that will be introduced in the following section, and $D_{\theta}(\mathbf{x}_t, t, cond)$ is designed to precondition the neural network with a t-dependent skip connection   as
\begin{equation}
D_{\theta}(\mathbf{x}_t , t,cond)=c_{\text {skip }}(t) \mathbf{x}_t +c_{\text {out }}(t) F_{\mathbf{\theta}}(\mathbf{x}_t, t,cond)
\end{equation}
where $F_{\mathbf{\theta}}$ is the network to be trained whose architecture can be flexibly chosen. For instance, the architectures of WaveNet  \cite{liu2022diffsinger}\cite{oord2016wavenet} or U-Net \cite{popov2021grad}\cite{ronneberger2015u} can be selected to construct $F_{\mathbf{\theta}}$. The 
 $c_{\text {skip }}(t)$ and $c_{\text {out }}(t)$ are used to modulate the skip connection and scale  the magnitudes of $F_\theta$, which can be given by \cite{song2023consistency}
\begin{equation}
c_{\mathrm{skip}}(t)=\frac{\sigma_{\mathrm{data}}^{2}}{(t-\epsilon)^{2}+\sigma_{\mathrm{data}}^{2}},~~~~c_{\mathrm{out}}(t)=\frac{\sigma_{\mathrm{data}}(t-\epsilon)}{\sqrt{\sigma_{\mathrm{data}}^{2}+t^{2}}}, 
\label{eq_skip_out}
\end{equation}
 where $\sigma_{\mathrm{data}}=0.5$ is used to balance the ratio between $c_{\text {skip }} $ and $c_{\text {out }} $ 
 and $\epsilon = 0.002$ as the smallest time instant during sampling. The first reason for choosing the above formulation is it can meet \eqref{s2} since $c_{\text {skip }}(\epsilon) =1$ and $c_{\text {out }}(\epsilon) = 0$. The second reason is that both scaling factors can help the predicted results of $F_{\mathbf{\theta}}$ scale to the unit variance, which avoids the large variation in gradient magnitudes at different noise levels.

To train the $D_{\theta}$, the loss function can be formulated as 
\begin{equation}
\label{diffusion}
 \mathcal{L}_\theta =    ||  D_{\theta}(\mathbf{x}_t , t,cond) - \mathbf{x}_0||^2,
\end{equation}
which is a weighted $\mathcal{L}_2$ loss between the predicted mel-spectrogram $pred_{mel}$ and ground truth  mel-spectrogram $gt_{\textrm{mel}}$, and we also re-weight the loss function for different $t$  as the same as EDM \cite{karras2022elucidating}. 

Finally, the teacher model can be trained, and the synthesized mel-spectrogram can be sampled by    Algorithm \ref{a_teacher}. During the inference on the teacher model,  we first sample $\mathbf{x}_{N}$  from $\mathcal{N}(\mu,I)$, and then  
iterative the numerical ODE solver for $N$ Euler steps. 
\begin{algorithm}
\caption{Sampling procedure of the proposed teacher model}\label{a_teacher}
\begin{algorithmic}[1]
\Statex \textbf{Input:}  The denoiser function $D_\theta$; the prior mel-spectrogram $\mu$; a set of time points $t_{i\in\{0,\ldots,N\}}$
\State Sample $\mathbf{x}_{N} \sim \mathcal{N}(\mu,I)$
\State $\mathbf{x}_{N} = t_{N} \mathbf{x}_{N}$
\State \textbf{for} $i = N$ \textbf{to} $1$ \textbf{do}
 
\State  \hskip1.0em $d_i \leftarrow  ( \mathbf{x}_{i} - D_{\theta}(\mathbf{x}_{i},t_{i},\mu)) / t_i$
\State \hskip1.0em  $\mathbf{x}_{i-1} \leftarrow  \mathbf{x}_{i} + (t_{i}-t_{i-1})d_i$
 
\State \textbf{end for}
\State $\mathbf{x} \leftarrow \mathbf{x}_{0}$
\Statex \textbf{Output:} $\mathbf{x}$
 
\end{algorithmic}
\end{algorithm}

\begin{algorithm}
\caption{Sampling procedure of the proposed  method}\label{a_como}
\begin{algorithmic}[1]
\Statex \textbf{Input:}  The denoiser function $D_\theta$; the prior mel-spectrogram $\mu$; a set of time points $t_{i\in\{0,\ldots,N\}}$
\State Sample $\mathbf{x}_{N} \sim \mathcal{N}(\mu,I)$
\State $\mathbf{x}_{N} = t_{N} \mathbf{x}_{N}$
\State  $\mathbf{x} \leftarrow D_{\theta}(\mathbf{x}_{N},t_{N},\mu)$ 
\State  \textbf{if}  one-step synthesis
\State \hskip1.0em \textbf{Output:} $\mathbf{x}$
\State  \textbf{else}            multi-step synthesis
\State \hskip1.0em \textbf{for} $i = N-1$ \textbf{to} $1$ \textbf{do}                
\State \hskip1.0em  \hskip1.0em  Sample $\textbf{z} \sim \mathcal{N}(\mu,I)$
\State \hskip1.0em \hskip1.0em  $\mathbf{x}_{i} \leftarrow  \mathbf{x} +\sqrt{t_{i}^{2}-\epsilon^{2}}{\bf z} $
\State \hskip1.0em \hskip1.0em  $\mathbf{x} \leftarrow D_{\theta}(\mathbf{x}_{i},t_{i},\mu)$                        
\State \hskip1.0em \textbf{end for}
\Statex \hskip1.0em \textbf{Output:} $\mathbf{x}$
 
\end{algorithmic}
\end{algorithm}

\subsection{Consistency Distillation}
A one-step diffusion sampling-based model is further trained from the teacher model based on consistency distillation, resulting in the proposed CoMoSpeech. Now we re-examine the constraints defined in \eqref{s1} and \eqref{s2}. We note that given the choice of $c_{\text {skip }}(t)$ and $c_{\text {out }}(t)$ in \eqref{eq_skip_out}, the denoiser $D_{\theta}$ in the proposed teacher model already satisfies \eqref{s2}, therefore, the remaining training objective is to fulfill the property in \eqref{s1}.

Inspired by \cite{song2023consistency}, we utilize the momentum-based distillation to train the proposed CoMoSpeech. The consistency distillation loss is defined as
\begin{equation}
\label{cd}
\mathcal{L}_\theta =||D_{\theta}(\mathbf{x}_{i+1} ,t_{i+1},cond)-D_{{\theta}^{-}}(\hat{\mathbf{x}}^{\mathbf{\phi}}_{i}, t_i,cond)||^2,
\end{equation}
where $\theta$ and $\theta^{-}$ are initialized weights of CoMoSpeech inherited from the teacher model, $\mathbf{\phi}$ is the fixed ODE solver from the teacher model in section \ref{teacher}, and $i$ is a step-index uniformly sampled from the total ODE steps from $N$ to 1. $\hat{\mathbf{x}}^{\mathbf{\phi}}_{i}$ is estimated from $x_{i+1}$ and the ODE solver $\mathbf{\phi}$. During training,    the weight $\theta$ directly optimize by the loss function, and $\theta^{-}$ is recursively updated by 
\begin{align}
    \theta^{-}  \leftarrow \mathrm{{stopgrad}} (\alpha\theta^{-}+(1-\alpha)\theta),
\end{align}
where $\alpha$ is a  momentum coefficient empirically set as 0.95.

\begin{figure}[!t]
      \centering
      \includegraphics[scale=0.55]{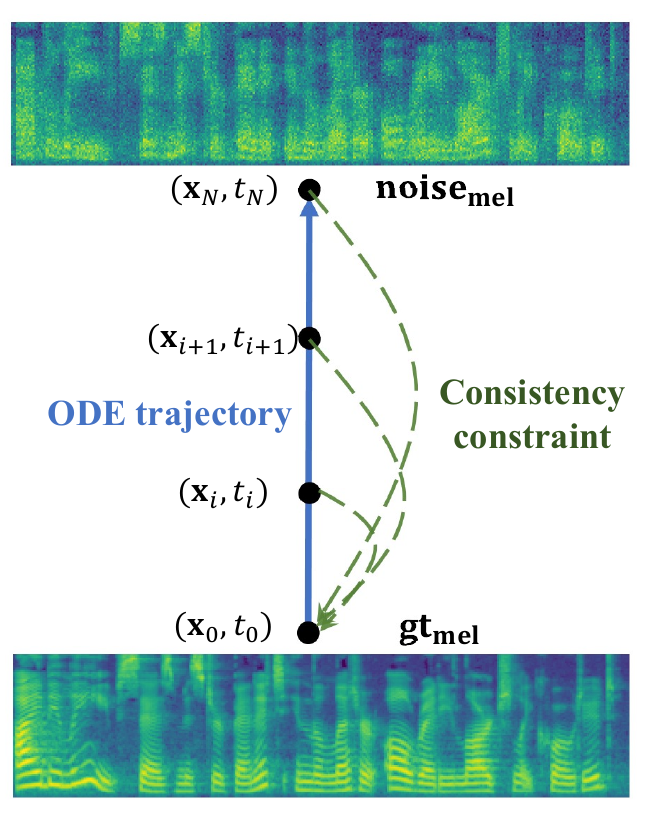}
      \caption{ An illustration of consistency property. A function with consistency property maps any points on the  ODE trajectory to the original data.}
      \label{fig2}
\end{figure}

After distillation, the consistency property can be exploited so that the original data point $\mathbf{x}_0$ can be transformed from any point $\mathbf{x}_t$ on the ODE trajectory as shown in Figure \ref{fig2}. Therefore,
we can directly generate the target sample  from the distribution    $\mathbf{x}_{N}  $   at the step $t_N$, as:
\begin{equation}
\label{one}
mel_{\textrm{pred}} = D_{\theta}(\mathbf{x}_{N},t_{N},cond).
\end{equation}
Therefore, one-step mel-spectrogram generation can be achieved. In addition, multi-step synthesis can be conducted  by Algorithm \ref{a_como} as a trade-off between audio quality and sampling speed, similar to other stochastic samplers.

\subsection{Conditional Input}
\label{ci}
A remaining problem in the framework shown in Figure~\ref{fig1} is how to obtain the conditional input $cond$, which will be used throughout the algorithm design. 
A well-designed speech synthesizer is expected to perform well not only on reading speech synthesis (TTS) but also on other more complicated tasks, such as SVS which additionally produces highly dynamic melodies. In producing the conditional inputs, both TTS and SVS tasks are considered to examine the proposed framework's effectiveness comprehensively. 

Concretely, we adopt the phoneme as the basic input for TTS and SVS. Then, a simple lookup table is used for embedding the phoneme feature. Additionally, for the SVS task, we add a music score that specifies the note levels time-aligned to the phonemes. For note feature extraction, we use the embedding method for both categorical feature note pitch and slur indicator and rely on a linear layer for continuous feature note duration.

Summing all the feature sequences together, we utilize the encoder structure and variance adaptor in FastSpeech \cite{ren2019fastspeech}. Specifically, $N$ feed-forward transformer blocks (FFT blocks) are stacked to extract the phoneme hidden sequence. A duration predictor is used to estimate the duration of each phoneme $d_{\textrm{pred}}$, and the corresponding loss function is expressed as
\begin{equation}
\label{duration}
\mathcal{L}_{\textrm{duration}} =||\log(d_{\textrm{pred}})-\log(d_{\mathrm{gt}})||^2
\end{equation}
where $d_{\textrm{gt}}$ indicates the ground-truth phoneme duration. 

Further, the length regulator projects the phoneme hidden sequence into the hidden sequence in the mel-spectrogram domain, with the phoneme duration denoted as $hidden_{\textrm{mel}}$. Then, the prior mel-spectrogram $\mu$ is predicted using the $hidden_{\textrm{mel}}$ with prior loss function as 
\begin{equation}
\label{prior}
\mathcal{L}_{\textrm{prior}} =||\mu-gt_{\textrm{mel}}||^2.
\end{equation}
We follow the encoder part and the prior mel-spectrogram setting in Grad-TTS \cite{popov2021grad}. As shown in the bottom left part of Figure \ref{fig1}, since the expanded features belonging to the same phoneme in $hidden_{\textrm{mel}}$ are repeated, the predicted $prior_{\textrm{mel}}$
can only roughly approximate the  time-frequency structure of $gt_{\textrm{mel}}$ based on the phoneme sequence.  The details of the mel-spectrogram are modeled by the diffusion model

For the neural network and conditional inputs in the denoiser, we investigated different combinations and finally followed the same setting in the previous work for a fair comparison. a) Diffsinger: the WaveNet architecture\cite{oord2016wavenet} and $hidden_{\textrm{mel}}$ as feature $cond$ in \eqref{one} for SVS and b) Grad-TTS: U-Net architecture \cite{ronneberger2015u} and $\mu_{\textrm{mel}}$ as $cond$ for TTS.

\subsection{Training procedure}
The whole process can be summarised as two stages which are the training of the teacher model and the consistency distillation.

As for the training of the teacher model, the loss term can consist of three parts which are duration loss (in \eqref{duration}), prior loss (in \eqref{prior}), and denoising loss (in \eqref{diffusion}).  
These three losses are summed up together without any extra weight. The objective of this stage is to build a  speech synthesis system that can generate high-quality audios with multi-step synthesis and have the potential for further consistency distillation.

The second stage is consistency distillation. There is only one loss function as defined by \eqref{cd}, which helps the model learn the consistency property. The  parameters are initialized from the teacher model. During training, the parameters of the encoder are fixed, which means only the weight in the denoiser is updated. After distillation, high-quality recordings with one-step synthesis \eqref{one} can be achieved.

\section{Experiments}
 
 \begin{table*}[t]
 \centering
 \setlength{\tabcolsep}{5.3mm}{\begin{tabular}{@{}lccccc@{}}
\toprule
METHOD           &NFE                      & RTF (↓)       &FD (↓)           & MOS (↑)                \\ \toprule
GT                   &   /      &  /  &    /   &  4.778\\
GT(Mel+HiFi-GAN)      &   /                        & /     &  0.282              &   4.590              \\  
FastSpeech 2 \cite{ren2020fastspeech}                          &  1            &   0.0017\     & 10.48   &4.034         \\
DiffGAN-TTS   \cite{liu2022diffgan} & 4    &0.0084    &  8.310   &3.889  \\
ProDiff \cite{huang2022prodiff}   &4    &0.0097    &3.503   &3.374  \\
DiffSpeech \cite{liu2022diffsinger} &71  &  0.1030 & 2.349  &4.103 \\
Grad-TTS \cite{popov2021grad} & 50  & 0.1694 &          1.882   &4.487 \\  \midrule

Teacher   &  50           & 0.1824 &   \textbf{0.748} & \textbf{4.538}  \\
\textbf{CoMoSpeech}            & \textbf{1}          & \textbf{0.0058}   & 0.774  &    4.239      \\ \bottomrule
 
\end{tabular}}
\caption{Evaluation results on LJSpeech for TTS.}
\label{t2}
\end{table*}

\begin{table*}[ht]
 \centering
 \setlength{\tabcolsep}{5.3mm}{
\begin{tabular}{@{}lccccc@{}}
\toprule
METHOD           &NFE                      & RTF (↓)       &FD (↓)            & MOS (↑)                \\ \toprule
GT                   &   /      &  /  &    /   &    4.675 \\
GT(Mel+HiFi-GAN)      &   /                        & /     &  0.882              &     4.588                  \\
 
FFTSinger   \cite{blaauw2020sequence}                          &  1            & 0.0032    & 7.867    &  2.769       \\
HIFiSinger  \cite{chen2020hifisinger}                     &1                      &   0.0034     & 6.340        &  3.156    \\  
DiffSinger \cite{liu2022diffsinger} A version & 60  & 0.1338 &  3.466    & 3.506       \\  
DiffSinger \cite{liu2022diffsinger} B version  &100  & 0.2198 & 3.618  &  3.531 \\ \midrule
CoMoSVS-teacher  &  50          & 0.1282& \textbf{3.162 } & \textbf{4.050}  \\
\textbf{CoMoSVS}            & \textbf{1}          &  \textbf{0.0048}   &   3.571    &3.794  \\ \bottomrule
 
\end{tabular}}
\caption{Evaluation Results on Opencpop for SVS.}
\label{t1}
\end{table*}

To evaluate the performance of the proposed CoMoSpeech, we conduct experiments on both TTS and SVS. 
\subsection{Experimental Setup}
\subsubsection{Data and Preprocessing}
We adopt the public LJSpeech \cite{ljspeech17} as the TTS dataset, which includes around $24$ hours of English female voice recordings sampled at $22.05$~kHz. Similar to \cite{ren2019fastspeech} \cite{chen2022resgrad}, we split the dataset into three sets: $12,228$ samples for training, $349$ samples (with document title LJ003) for validation, and $523$ samples (with document title LJ001 and LJ002) for testing. Following the common practice in \cite{ren2020fastspeech}\cite{huang2022prodiff} for TTS, we extract the 80-bin mel-spectrogram with the frame size of $1024$ and hop size of $256$.
 
For the SVS task, we use the Opencpop dataset \cite{wang2022opencpop} containing 100 Chinese pop songs which are split into $3,756$ utterances with a total duration of around $5.2$ hours. All recordings are from a single female singer and labeled with aligned phoneme and MIDI-pitch sequences. We follow the official train/test split \cite{wang2022opencpop}, i.e., $95$ songs and $5$ songs for training and evaluation, respectively.
Same as the setting in \cite{chen2020hifisinger}\cite{liu2022diffsinger} for SVS, the recordings are resampled at $24$kHz rates with $16$-bit precision, and the $80$-bin mel-spectrogram is extracted with a frame size of $512$ and hop size of $128$.

\subsubsection{Implementation Details}

For TTS, for a fair comparison, the encoder and duration predictor are exactly the same as those in Grad-TTS \cite{popov2021grad}. The encoder contains $6$ feed-forward transformer (FFT) blocks \cite{ren2019fastspeech}, and The hidden channel is set to $192$. The duration predictor uses two convolutional layers for prediction. 
Both the teacher model and CoMoSpeech are trained for $1.7$ million iterations on a single NVIDIA A100 GPU with a batch size of $16$. The Adam optimizer \cite{kingma2014adam} is adopted with the learning rate 1e-4.

For SVS, we adopt almost the same architecture as in TTS with different hyperparameters. The encoder adopts $4$ FFT blocks, and we set the hidden channel to $256$ in the encoder. The duration predictor consists of $5$ convolutional layers to estimate the duration. The teacher model of SVS and CoMoSpeech are trained on a single GPU for $250$k steps with the AdamW \cite{loshchilov2017decoupled} optimizer. The initial learning rate is 1e-3, and the exponential decay strategy  with a decreasing factor of $0.5$ every $50$k steps is adopted.

\subsubsection{Evaluation Metrics}
we conduct both objective  and  subjective evaluations to measure the sample quality (MOS \& FD) and the model inference speed (RTF \& NFE):
\begin{itemize}

\item
MOS (mean opinion score) \cite{chu2006objective}  is  used to measure the perceived quality of the synthesized audio, which is obtained by presenting 10 listeners with the test set and asking them to rate the quality of the synthesized audio on a scale of 1 to 5. 
\item
FD (frechet distance)\footnote{https://github.com/haoheliu/audioldm\_eval} is similar to the frechet inception distance \cite{heusel2017gans} in image generation. We use  frechet distance \cite{liu2023audioldm} in audio to measure the similarity between generated samples and target samples utilizing  the large-scale pretrained audio neural networks PANNs \cite{kong2020panns}.
\item
RTF (real-time factor)  determines how quickly the system can synthesize audio in real-time applications. It is defined as the ratio between the total time a speech system takes to synthesize a given amount of audio and the duration of that audio. In addition, all experiments for RTF are implemented on a single NVIDIA A100 GPU. 
\item

NFE (number of  function evaluations) measures the computational cost, which refers to the total number of times the denoiser function is evaluated during the generation process.

\end{itemize}

\subsection{Performances  on Text-to-Speech}
We compare the above four metrics of the samples generated by the teacher model and CoMoSpeech with the following systems:

\begin{itemize}
\item GT, the ground truth recordings.
\item GT (Mel+HiFi-GAN), using ground-truth mel-spectrogram to synthesize waveform with HiFi-GAN vocoder \cite{kong2020hifi}.
\item FastSpeech 2 \cite{ren2020fastspeech}, synthesizing high-quality speech at a fast speed with FFT blocks and variance adaptor. 

\item DiffGAN-TTS \cite{liu2022diffgan} \footnote{https://github.com/keonlee9420/DiffGAN-TTS}, applying an adversarially-trained model to approximate the denoising function for efficient speech synthesis.

\item ProDiff  \cite{huang2022prodiff} \footnote{https://github.com/Rongjiehuang/ProDiff}, directly adopting progressive distillation \cite{salimans2022progressive} to TTS for fast generation speed.

\item DiffSpeech \cite{liu2022diffsinger} \footnote{https://github.com/MoonInTheRiver/DiffSinger/blob/master/docs/README-TTS.md},  using an auxiliary acoustic model to generate mel-spectrogram and injects K steps noise to a noisy  mel-spectrogram. Then, the  mel-spectrogram is generated from the noisy  mel-spectrogram by DDPM iteratively. 

\item Grad-TTS \cite{popov2021grad} \footnote{https://github.com/huawei-noah/Speech-Backbones/tree/main/Grad-TTS}, using stochastic differential equation modelling for the mel-spectrogram and use corresponding ODE solver for audio generation.

\end{itemize}

 \begin{figure*}[!h]
      \centering
      \includegraphics[width=130mm]{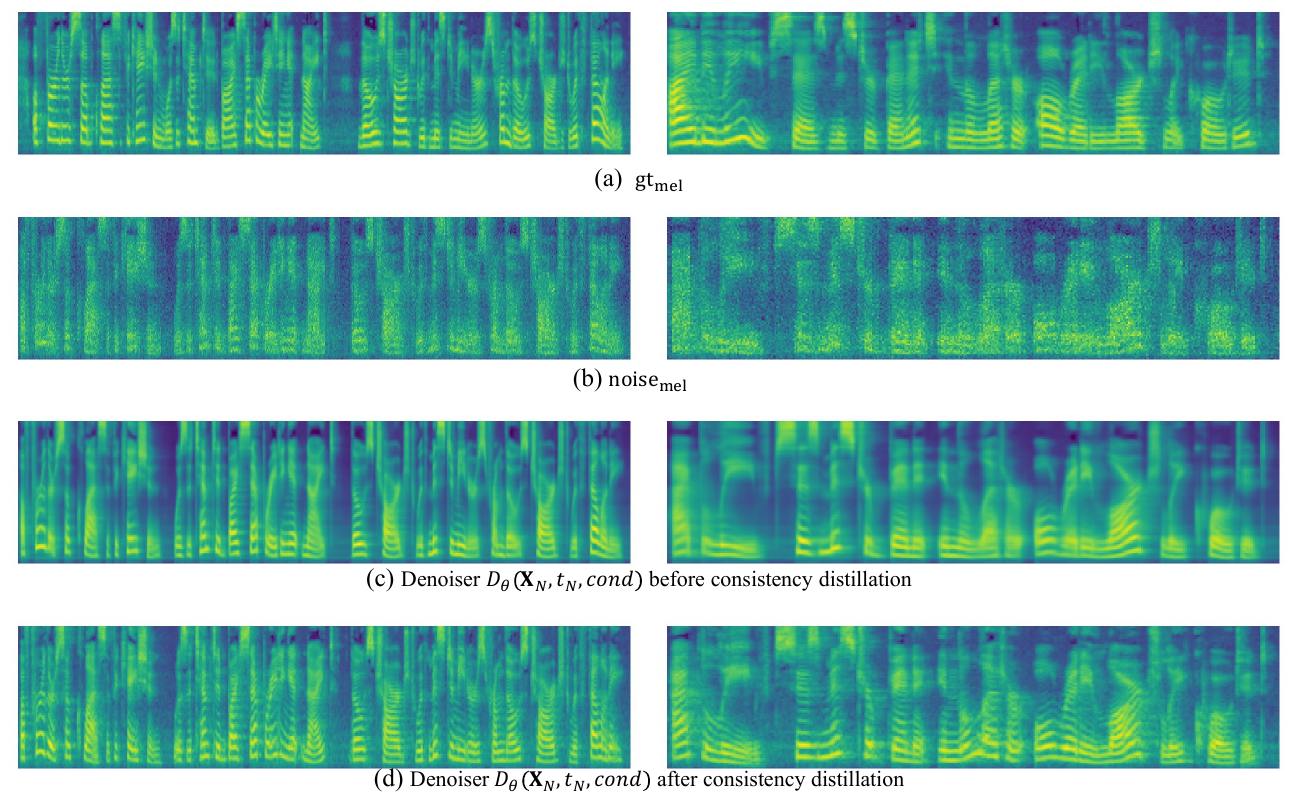}
      \caption{ Effect of  consistency distillation: Compared to the denoiser of teacher model before consistency distillation, our CoMoSpeech can generate a high-quality mel-spectrogram instead of an over-smoothed mel-spectrogram by calling the denoiser function only one time.}
      \label{fig3}
\end{figure*}

The evaluation results of TTS are shown in Table \ref{t2}. For audio quality, our teacher model achieved the highest MOS and Grad-TTS ranked second  because our teacher model is based on the design of Grad-TTS, but we adopt better choices on drift and diffusion coefficients in SDE. The proposed CoMoSpeech takes 3rd place among all methods, but it is substantially better than other fast-inference methods ProDiff, DiffGAN-TTS and Fastspeech2. This demonstrates the effectiveness of the consistency distillation and the effectiveness of teacher model selection. In addition, we also observe that our teacher model and CoMoSpeech achieve the best frechet distance scores among all methods, further demonstrating the proposed methods' superior performance on modeling data distribution.
 
Regarding inference speed, while Fastspeech2 obviously achieves the best,  our CoMoSpeech also yields a very low RTF, and is faster than all other baselines. Compared with the diffusion-based methods involving a large number of iterations including DiffSpeech, Grad-TTS and our teacher model, our method achieves about 50 times faster with similar or even better audio quality. In addition, our CoMoSpeech also achieves faster speed and better quality than methods for speeding up diffusion sampling, i.e., DiffGAN-TTS and ProDiff.

\subsection{Performances on Singing Voice Synthesis}
To further examine the modeling capability of our methods, we compare the proposed SVS-version models, teacher-svs and CoMoSpeech-svs, with several baselines on SVS, and the baselines include:
\begin{itemize}
\item GT, the ground truth recordings.
\item GT(Mel+HiFi-GAN), synthesizing song samples using HiFi-GAN \cite{kong2020hifi} vocoder with ground truth mel-spectrogram inputs.
\item FFTSinger \cite{blaauw2020sequence}, adopting FFT blocks to predict mel-spectrogram, and using the HiFi-GAN vocoder to synthesize audio;
\item HiFiSinger \cite{chen2020hifisinger}, using a novel sub-frequency GAN (SF-GAN)  to generate the mel-spectrogram. Since our aim is to compare the acoustic model, we modify the original vocoder to HiFi-GAN, being the same as other methods.
\item DiffSinger \cite{liu2022diffsinger},  using DDPM to generate the mel-spectrogram from noisy  mel-spectrogram. There are two versions for generating noisy mel-spectrogram where A version \footnote{Version A: https://github.com/MoonInTheRiver/DiffSinger/blob/master/docs/README-SVS-opencpop-cascade.md} using a auxiliary acoustic model to generate mel-spectrogram and injecting K steps noise to a noisy  mel-spectrogram, and B version  \footnote{Version B: https://github.com/MoonInTheRiver/DiffSinger/blob/master/docs/README-SVS-opencpop-e2e.md} directly generating the noisy mel-spectrogram from the Gaussian noise.
 
\end{itemize}

The results of SVS are shown in Table \ref{t1}.  As for audio quality, it can be seen that our CoMoSpeech and other diffusion model based methods can significantly surpass all non-iterative methods including FFTSinger and HIFiSinger on frechet distance and mean opinion score. Among diffusion models, our teacher model achieves the best performance, and our student model CoMoSpeech has a close performance to it. For the inference speed, with one-step inference, the proposed CoMoSpeech could maintain a speed similar to non-iterative methods and significantly outperform other diffusion model based methods.  

In addition, we also compare the results between speaking voice and singing voice synthesis. Based on two methods DiffSinger and DiffSpeech, which are basically the same, we can observe that the singing voice has a greater FD than the speaking voice, indicating that it is more difficult to model the data. However, the proposed teacher model and CoMoSpeech still achieve the best performances on audio quality and inference speed, respectively. This shows the capability of CoMoSpeech for speech synthesis beyond speaking voices. In addition, we can observe that our CoMoSpeech-svs is faster than CoMoSpeech because the denoiser function in SVS follows the WaveNet architecture which is faster than U-Net architecture in TTS. This observation inspires us that if a more efficient denoiser function that runs faster than the decoder in FastSpeech2 can be designed, we can make CoMoSpeech even faster than non-iterative methods in future work.

\begin{table}[]
\centering
\begin{tabular}{c|cc}
\hline
& \multicolumn{2}{c}{Frechet Distance (↓) }                          \\ \cline{2-3} 
 NFE    & \multicolumn{1}{c|}{Teacher model} & CoMoSpeech \\ \cline{2-3} 
                         \hline
1                            & 7.526            &    0.774         \\
2                            & 4.558     &     0.762        \\
4                             & 2.477         &    0.784         \\
10                             & 1.197      &       \textbf{ 0.725 }   \\
50                             &  \textbf{0.748  }   &      0.850    \\  \hline
\end{tabular}
\caption{Comparison  between CoMoSpeech and its teacher model with different sampling steps for TTS.}
\label{t3:tts}
\end{table}

\begin{table}[]
\centering
\begin{tabular}{c|cc}
\hline
& \multicolumn{2}{c}{Frechet Distance (↓) }                          \\ \cline{2-3} 
 NFE    & \multicolumn{1}{c|}{Teacher model} & CoMoSpeech \\ \cline{2-3} 
                         \hline
1                            &7.786         &    3.571        \\
2                            &7.219        &       3.520    \\
4                             & 4.932        &      \textbf{3.433}       \\
10                             & 3.937      &     3.658        \\
50                             &  \textbf{3.162}     &   3.732       \\ \hline
\end{tabular}
\caption{Comparison  between CoMoSpeech and its teacher model with different sampling steps for SVS.}
\label{t4:svs}
\end{table}



\subsection{Ablation Studies of Consistency Distillation}
In this part, we will show the importance of consistency distillation. As shown in Figure \ref{fig3}, we visualize the differences before and after consistency distillation in the results,in other words, the teacher model and our CoMoSpeech. 
At $t_N$ steps, the denoiser function before distillation points to a smooth mel-spectrogram, indicating a great distance between ground truth mel-spectrogram. However, we can observe that the results after distillation significantly improve the performances by enriching many details, resulting in natural and expressive sounds.  

In Table~\ref{t3:tts} and Table~\ref{t4:svs}, we also conduct experiments using the frechet distance metric to further demonstrate the effectiveness of consistency distillation. For teacher models for both TTS and SVS tasks, the frechet distance decrease when the iteration steps increase. This  trade-off property   between inference speed and sample quality has also been observed in other diffusion model methods. Surprisingly, we can find that our CoMoSpeech can achieve nearly the best performance in one step, and the best performance can be achieved at 4 and 10 steps on TTS and SVS, respectively. However, the trade-off property seems to disappear after $10$ steps. The issue that the model performance improves in a few
sampling steps and then declines slightly as the number of steps increases is called  the "sampling drift" challenge \cite{daras2023consistent} \cite{ji2023ddp} \cite{chen2022sampling} \cite{saxena2023monocular}. We will leave the exploration to future work.

\section{Conclusions and Future Work}
In this paper, we propose CoMoSpeech, a one-step acoustic model for speech synthesis based on the consistency model. With different conditional inputs, our CoMoSpeech can generate high-quality speech or singing voice by transforming the noise mel-spectrogram into the predicted mel-spectrogram in a single step. 

However, there are still some limitations to our method. Since our CoMoSpeech needs to be distilled from a teacher model for better performance, this makes the pipeline of constructing a speech synthesis system more complicated. Therefore, how to directly train the CoMoSpeech without distillation of the teacher model is our next step to investigate. In addition, we show the capability of CoMoSpeech on the SVS task. Even though CoMoSpeech achieves the best result among all the methods, it still has a significant gap between the ground truth recording.

\section*{Acknowledgments}
The research was supported by the Theme-based Research Scheme (T45-205/21-N) and Early Career Scheme (ECS-HKUST22201322), Research Grants Council of Hong Kong.

\bibliographystyle{plainnat}
\bibliography{sample-base}


\end{document}